\documentclass[reprint, showpacs, preprintnumbers, amsmath,amssymb,  prb,
twocolumn ]{revtex4-1}%
\usepackage{epsf}
\usepackage{graphicx}
\usepackage{color}
\usepackage{amsmath}
\usepackage{amsfonts}
\usepackage{amssymb}
\usepackage{dcolumn}
\usepackage{txfonts}%
\providecommand{\U}[1]{\protect\rule{.1in}{.1in}}

\begin{document}
\title{Magnetization damping in noncollinear spin valves with antiferromagnetic
interlayer couplings}
\date{\today}
\author{Takahiro Chiba$^{1}$, Gerrit E. W. Bauer$^{1,2,3}$, and Saburo Takahashi$^{1}$}
\affiliation{$^{1}$ Institute for Materials Research, Tohoku University, Sendai, Miyagi
980-8577, Japan}
\affiliation{$^{2}$ WPI-AIMR, Tohoku University, Sendai, Miyagi 980-8577, Japan}
\affiliation{$^{3}$ Kavli Institute of NanoScience, Delft University of Technology,
Lorentzweg 1, 2628 CJ Delft, The Netherlands}

\begin{abstract}
We study the magnetic damping in the simplest of synthetic antiferromagnets,
i.e. antiferromagnetically exchange-coupled spin valves in which applied
magnetic fields tune the magnetic configuration to become noncollinear. We
formulate the dynamic exchange of spin currents in a noncollinear texture
based on the spin-diffusion theory with quantum mechanical boundary conditions
at the ferrromagnet$|$normal-metal interfaces and derive the
Landau-Lifshitz-Gilbert equations coupled by the static interlayer non-local
and the dynamic exchange interactions. We predict non-collinearity-induced
additional damping that can be sensitively modulated by an applied magnetic
field. The theoretical results compare favorably with published experiments.

\end{abstract}
\maketitle
\affiliation{Institute for Materials Research, Tohoku University, Sendai 980-8577, Japan}
\affiliation{Institute for Materials Research, Tohoku University, Sendai 980-8577, Japan}
\affiliation{WPI-AIMR, Tohoku University, Sendai 980-8577, Japan}
\affiliation{Kavli Institute of NanoScience, Delft University of Technology, Lorentzweg 1,
2628 CJ Delft, The Netherlands}
\affiliation{Institute for Materials Research, Tohoku University, Sendai 980-8577, Japan}



\section{Introduction}

Antiferromagnets (AFMs) boast many of the functionalities of ferromagnets (FM)
that are useful in spintronic circuits and devices: Anisotropic
magnetoresistance (AMR),\cite{Marti14} tunneling anisotropic magnetoresistance
(TAMR),\cite{Park11} current-induced spin transfer
torque,\cite{Haney07,Sharma07,Urazhdin07,Wang08,Hals11,Gomonay12} and spin
current transmission\cite{Hahn14,Wang14,Moriyama14} have all been found in or
with AFMs. This is of interest because AFMs have additional features
potentially attractive for applications. In AFMs the total magnetic moment is
(almost) completely compensated on an atomic length scale. The AFM order
parameter is, hence, robust against perturbations such as external magnetic
fields and do not generate stray fields themselves either. A spintronic
technology based on AFM elements is therefore very
attractive.\cite{Duine11,MacDonald11} Drawbacks are the difficulty to
\textit{control} AFMs by magnetic fields and much higher (THz) resonance
frequencies,\cite{Kimel05,Satoh10,Wienholdt12} which are difficult to match
with conventional electronic circuits. Man-made magnetic multilayers in which
the layer magnetizations in the ground state is ordered in an antiparallel
fashion, \cite{Grunberg86} i.e. so-called synthetic antiferromagnets, do not
suffer from this drawback and have therefore been a fruitful laboratory to
study and modulate antiferromagnetic couplings and its
consequences,\cite{Grunberg07} but also found applications as magnetic field
sensors.\cite{Parkin91} Transport in these multilayers including the giant
magnetoresistance (GMR)\cite{Baibich88,Binasch89} are now well understood in
terms of spin and charge diffusive transport. Current-induced magnetization
switching in F$|$N$|$F spin valves and tunnel junctions,\cite{Brataas12} has
been a game-changer for devices such as magnetic random access memories
(MRAM).\cite{Chen10} A key parameter of magnetization dynamics is the magnetic
damping; a small damping lowers the threshold of current-driven magnetization
switching,\cite{Ralph08} whereas a large damping suppresses \textquotedblleft
ringing\textquotedblright\ of the switched magnetization.\cite{Carey08}

Magnetization dynamics in multilayers generates \textquotedblleft spin
pumping\textquotedblright, i.e. spin current injection from the ferromagnet
into metallic contacts. It is associated with a loss of angular momentum and
an additional interface-related magnetization
damping.\cite{Tserkovnyak02,Tserkovnyak05} In spin valves, the additional
damping is suppressed when the two magnetizations precess in-phase, while it
is enhanced for a phase difference of $\pi$
(out-of-phase).\cite{Tserkovnyak05,Heinrich03, Takahashi14,Tanaka14} This
phenomenon is explained in terms of a \textquotedblleft dynamic exchange
interaction\textquotedblright, i.e. the mutual exchange of non-equilibrium
spin currents, which should be distinguished from (but coexists with) the
oscillating equilibrium exchange-coupling mediated by the
Ruderman-Kittel-Kasuya-Yosida (RKKY) interaction. The equilibrium coupling is
suppressed when the spacer thickness is thicker than the elastic mean-free
path,\cite{Zhang94,Belmeguenai07} while the dynamic coupling is effective on
the scale of the usually much larger spin-flip diffusion length.

Antiparallel spin valves provide a unique opportunity to study and control the
dynamic exchange interaction between ferromagnets through a metallic
interlayer for tunable magnetic configurations.\cite{Stamps94,Hans14} An
originally antiparallel configuration is forced by relatively weak external
magnetic fields into a non-collinear configuration with a ferromagnetic
component. Ferromagnetic resonance (FMR) and Brillouin light scattering (BLS)
are two useful experimental methods to investigate the nature and magnitude of
exchange interactions and magnetic damping in multilayers.\cite{Demokritov01}
Both methods observe two resonances, i.e. acoustic (A) and optical (O) modes,
which are characterized by their frequencies and
linewidths.\cite{Cochran90,Kuanr02}

Timopheev\textit{ et al.} observed an effect of the interlayer RKKY
coupling on the FMR and found the linewidth to be affected by the dynamic
exchange coupling in spin valves with one layer fixed by the exchange-bias of
an inert AFM substrate.\cite{Timopheev14} They measured the FMR spectrum of the free layer by
tuning the interlayer coupling (thickness) and reported a broadening of the
linewidth by the dynamic exchange interaction. Taniguchi\textit{ et al.}
addressed theoretically the enhancement of the Gilbert damping constant due to
spin pumping in noncollinear F$|$N$|$F trilayer systems, in which one of the
magnetizations is excited by FMR while the other is off-resonant, but adopt a
role as spin sink.\cite{Taniguchi07}
\textbf{ }The dynamics of coupled spin valves in which both layer
magnetizations are free to move has been computed by one of
us\cite{Takahashi14} and by Skarsv\r{a}g \textit{ et al.}\cite{Hans14,Hans14SW} but only
for collinear (parallel and antiparallel) configurations. Current-induced
high-frequency oscillations without applied magnetic field in
ferromagnetically coupled spin valves has been predicted.\cite{Zhou13}

In the present paper, we model the magnetization dynamics of the simplest of
synthetic antiferromagnets, i.e. the antiferromagnetically exchange-coupled
spin valve in which the (in-plane) ground state magnetizations are for certain
spacer thicknesses ordered in an antiparallel fashion by the RKKY interlayer
coupling.\cite{Layadi97} We focus on the coupled magnetization modes in
symmetric spin valves in which in contrast to previous studies, both
magnetizations are free to move. We include static magnetic fields in the film
plane that deform the antiparallel configuration into a canted one. Microwaves
with longitudinal and transverse polarizations with respect to an external
magnetic field then excite A and O resonance modes,
respectively.\cite{Krebs90,Zhang94,Kuanr03,Evans14,Gonzalez-Chavez13,Liu14} We
develop the theory for magnetization dynamics and damping based on the
Landau-Lifshitz-Gilbert equation with mutual pumping of spin currents and spin
transfer torques based on the spin diffusion model with quantum mechanical
boundary conditions.\cite{Tserkovnyak05,Chiba14,Chiba15} We
confirm\cite{Heinrich03,Hans14SW} that the additional damping of O modes is
larger than that of the A modes. We report that a noncollinear magnetization
configuration induces additional damping torques that to the best of our
knowledge have not been discussed in magnetic multilayers before.\cite{Yuan14} The external magnetic
field strongly affects the dynamics by modulating the phase of the dynamic
exchange interaction. We compute FMR linewidths as a function of applied
magnetic fields and find good agreement with experimental FMR\ spectra on spin
valves.\cite{Zhang94,Belmeguenai07} The dynamics of magnetic multilayers as
measured by ac spin transfer torque excitation\cite{Tanaka14} reveals a
relative broadening of the O modes linewidths that is well reproduced by our
spin valve model.

\begin{figure}[ptb]
\includegraphics[width=0.45\textwidth,angle=0]{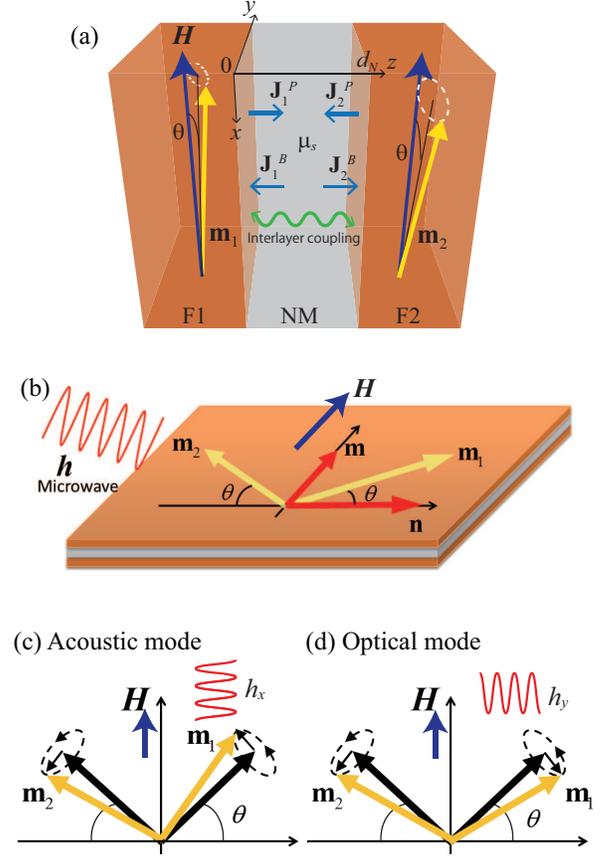}\label{fig:AFMR}%
\caption{(a) Sketch of the sample with interlayer exchange-couplings
illustrating the spin pumping and backflow currents. (b) Magnetic resonance in
an antiferromagnetically exchange-coupled spin valve with a normal-metal (NM)
film sandwiched by two ferromagnets (F1, F2) subject to a microwave magnetic
field $\mathbf{h}$. The magnetization vectors ($\mathbf{m}_{1}$,
$\mathbf{m}_{2}$) are tilted by an angle $\theta$ in a static in-plane
magnetic field $\mathbf{H}$ applied along the $y$-axis. The vectors
$\mathbf{m}$ and $\mathbf{n}$ represent the sum and difference of the two
layer magnetizations, respectively. (c) and (d): Precession-phase relations
for the acoustic and optical modes. }%
\label{sketch}%
\end{figure}

In Sec.~\ref{spin-diffusion} we present our model for noncollinear spin
valves based on spin-diffusion theory with quantum mechanical boundary
conditions. In Sec.~\ref{magnetization dynamics}, we consider the
magnetization dynamics in antiferromagnetically coupled noncollinear spin
valves as shown in Fig.~\ref{sketch}(b). We derive the linearized
magnetization dynamics, resonance frequencies, and lifetimes of the acoustic
and optical resonance modes in Sec.~\ref{result}. We discuss the role of
dynamic spin torques on noncollinear magnetization configurations in relation
to external magnetic field dependence of the linewidth. In
Sec.~\ref{experiment}, we compare the calculated microwave absorption and
linewidth with published experiments. We summarize the results and end with
the conclusions in Sec.~\ref{conclusion}.

\section{Spin diffusion transport model}
\label{spin-diffusion} 

We consider F1$|$N$|$F2 spin valves as shown in
Fig.~\ref{sketch}(a), in which the magnetizations $\mathbf{M}_{j}$ of the
ferromagnets F$j$ ($j=1,2$) are coupled by a antiparallel interlayer exchange
interaction and tilted towards the direction of an external magnetic field.
Applied microwaves with transverse polarizations with respect to an external
magnetic field cause dynamics and, via spin pumping, spin currents and
accumulations in the normal-metal (NM) spacer. The longitudinal component of
the spin accumulation diffuses into and generates spin accumulations in F that
we show to be small later, but disregard initially. Let us denote the pumped
spin current $\mathbf{J}_{j}^{\mathrm{P}}$, while $\mathbf{J}_{j}^{\mathrm{B}%
}$ is the diffusion (back-flow) spin current density induced by a spin
accumulation $\boldsymbol{\mu}_{sj}$ in NM, both at the interface F$j$,
with\cite{Tserkovnyak05,Jiao13}
\begin{subequations}
\begin{align}
\mathbf{J}_{j}^{P} &  =\frac{G_{r}}{e}\mathbf{m}_{j}\times{\hbar}\partial
_{t}\mathbf{m}_{j},\\
\mathbf{J}_{j}^{B} &  =\frac{G_{r}}{e}\left[  \left(  \mathbf{m}_{j}%
\cdot\boldsymbol{\mu}_{sj}\right)  \mathbf{m}_{j}-\boldsymbol{\mu}%
_{sj}\right]  ,\label{jsb}%
\end{align}
where $\mathbf{m}_{j}=\mathbf{M}_{j}/\left\vert \mathbf{M}_{j}\right\vert $ is
the unit vector along the magnetic moment of F$j$ ($j=1,2$). The spin current
through a FM$|$NM interface is governed by the complex spin-mixing conductance
(per unit area of the interface) $G^{\uparrow\downarrow}=G_{r}+iG_{i}%
$.\cite{Tserkovnyak05} The real component $G_{r}$ parameterized one vector
component of the transverse spin-currents pumped and absorbed by the
ferromagnets. The imaginary part $G_{i}$ can be interpreted as an effective
exchange field between magnetization and spin accumulation, which in the
absence of spin-orbit interaction is usually much smaller than the real part,
for conducting as well as insulting magnets.\cite{Jia11}

The diffusion spin-current density in NM reads
\end{subequations}
\begin{equation}
\mathbf{J}_{s,z}(z)=-\frac{\sigma}{2e}\partial_{z}\boldsymbol{\mu}_{s}(z),
\label{djs}%
\end{equation}
where $\sigma=\rho^{-1}$ is the electrical conductivity and $\boldsymbol{\mu
}_{s}(z)=\boldsymbol{A}e^{-z/\lambda}+\boldsymbol{B}e^{z/\lambda}$ the spin
accumulation vector that is a solution of the spin diffusion equation
$\partial_{z}^{2}\boldsymbol{\mu}_{s}=\boldsymbol{\mu}_{s}/\lambda^{2}$, where
$\lambda=\sqrt{D\tau_{\mathrm{sf}}}$ is the spin-diffusion length, $D$ the
diffusion constant, and $\tau_{\mathrm{sf}}$ the spin-flip relaxation time.
The vectors $\boldsymbol{A}$ and $\boldsymbol{B}$ are determined by the
boundary conditions at the F1$|$NM ($z=0$) and F2$|$NM ($z=d_{\mathrm{N}}$)
interfaces: $\mathbf{J}_{s,z}\left(  0\right)  =\mathbf{J}_{1}^{\mathrm{P}%
}+\mathbf{J}_{1}^{\mathrm{B}}\equiv\mathbf{J}_{s1}$ and $\mathbf{J}%
_{s,z}\left(  d_{\mathrm{N}}\right)  =-\mathbf{J}_{2}^{\mathrm{P}}%
-\mathbf{J}_{2}^{\mathrm{B}}\equiv-\mathbf{J}_{s2}$. The resulting spin
accumulation in N reads
\begin{equation}
\boldsymbol{\mu}_{s}(z)=\frac{2e\lambda\rho}{\sinh\left(  \frac{d_{\mathrm{N}%
}}{\lambda}\right)  }\left[  \mathbf{J}_{s1}\cosh\left(  \frac{z-d_{\mathrm{N}%
}}{\lambda}\right)  +\mathbf{J}_{s2}\cosh\left(  \frac{z}{\lambda}\right)
\right]  , \label{mus}%
\end{equation}
with interface spin currents
\begin{widetext}
\begin{subequations}
\begin{align}
\mathbf{J}_{s1}=&\frac{ \eta S}{1-\eta^2}\left[  \delta\mathbf{J}_{1}^{P} + \frac{\eta^2\left( \mathbf{m}_{2}\cdot\delta\mathbf{J}_{1}^{P}\right)}{1-\eta^2\left(\mathbf{m}_{1}\cdot\mathbf{m}_{2}\right)^2}
\mathbf{m}_{1}\times\left(\mathbf{m}_{1}\times\mathbf{m}_{2}\right)\right],\label{interjs-a}\\
\mathbf{J}_{s2}=&-\frac{ \eta S}{1-\eta^2}\left[  \delta\mathbf{J}_{2}^{P} + \frac{\eta^2\left( \mathbf{m}_{1}\cdot\delta\mathbf{J}_{2}^{P}\right)}{1-\eta^2\left(\mathbf{m}_{1}\cdot\mathbf{m}_{2}\right)^2}
\mathbf{m}_{2}\times\left(\mathbf{m}_{2}\times\mathbf{m}_{1}\right)\right]
.\label{interjs-b}%
\end{align}
\end{subequations}
\end{widetext}
Here
\begin{subequations}
\begin{align}
\delta\mathbf{J}_{1}^{P}  &  =\mathbf{J}_{1}^{P}+\eta\mathbf{m}_{1}%
\times(\mathbf{m}_{1}\times\mathbf{J}_{2}^{P}),\\
\delta\mathbf{J}_{2}^{P}  &  =\mathbf{J}_{2}^{P}+\eta\mathbf{m}_{2}%
\times(\mathbf{m}_{2}\times\mathbf{J}_{1}^{P}),
\end{align}
$S=\sinh(d_{\mathrm{N}}/\lambda)/g_{r}$ and $\eta=g_{r}/[\sinh(d_{\mathrm{N}%
}/\lambda)+g_{r}\cosh(d_{\mathrm{N}}/\lambda)]$ are the efficiency of the back
flow spin currents, and $g_{r}=2\lambda\rho G_{r}$ is dimensionless. The first
terms in Eqs.~(\ref{interjs-a}) and (\ref{interjs-b}) represent the mutual
pumping of spin currents while the second terms may be interpreted as a spin
current induced by the noncollinear magnetization configuration, including the
back flow from the NM interlayer.

\section{Magnetization dynamics with dynamic spin torques}
\label{magnetization dynamics}

We consider the magnetic resonance in the non-collinear spin valve shown in
Fig.~\ref{sketch}. The magnetization dynamics are described by the
Landau-Lifshitz-Gilbert (LLG) equation,
\end{subequations}
\begin{subequations}
\begin{align}
\partial_{t}\mathbf{m}_{1} &  =-\gamma\mathbf{m}_{1}\times\mathbf{H}%
_{\mathrm{eff}1}+\alpha_{0}\mathbf{m}_{1}\times\partial_{t}\mathbf{m}%
_{1}+\boldsymbol{\tau}_{1},\label{llg-a}\\
\partial_{t}\mathbf{m}_{2} &  =-\gamma\mathbf{m}_{2}\times\mathbf{H}%
_{\mathrm{eff}2}+\alpha_{0}\mathbf{m}_{2}\times\partial_{t}\mathbf{m}%
_{2}-\boldsymbol{\tau}_{2}.\label{llg-b}%
\end{align}
The first term in Eqs.~(\ref{llg-a}) and (\ref{llg-b}) represents the torque
induced by the effective magnetic field
\end{subequations}
\begin{equation}
\mathbf{H}_{\mathrm{eff}1(2)}=\mathbf{H}+\mathbf{h}(t)-4\pi M_{s}%
m_{1(2)z}\mathbf{\hat{z}}+\frac{J_{ex}}{M_{s}d_{\mathrm{F}}}\mathbf{m}_{2(1)},
\end{equation}
which consists of an in-plane applied magnetic field $\mathbf{H}$, a microwave
field $\mathbf{h}(t)$, and the demagnetization field $-4\pi M_{s}%
m_{1(2)z}\mathbf{\hat{z}}$ with saturation magnetization $M_{s}$. The
interlayer exchange field is $J_{ex}/(M_{s}d_{\mathrm{F}})\mathbf{m}_{2(1)}$ with areal
density of the interlayer exchange energy $J_{ex}<0$\thinspace(for
antiferromagnetic-coupling) and F layer thickness $d_{\mathrm{F}}$. The second term is
the Gilbert damping torque that governs the relaxation characterized by
$\alpha_{0i}$ towards an equilibrium direction. The third term,
$\boldsymbol{\tau}_{\mathrm{1(2)}}=\gamma\hbar/(2eM_{s}d_{\mathrm{F}})\mathbf{J}%
_{s1(2)}$, is the spin-transfer torque induced by the absorption of the
transverse spin currents of Eqs.~(\ref{interjs-a}) and (\ref{interjs-b}), and
$\gamma$ and $\alpha_{0}$ are the gyromagnetic ratio and the Gilbert damping
constant of the isolated ferromagnetic films, respectively. Some technical
details of the coupled LLG equations are discussed in Appendix~\ref{Appendix}.
Introducing the total magnetization direction $\mathbf{m}=(\mathbf{m}%
_{1}+\mathbf{m}_{2})/2$ and the difference vector $\mathbf{n}=(\mathbf{m}%
_{1}-\mathbf{m}_{2})/2$, the LLG equations can be written
\begin{subequations}
\begin{align}
\partial_{t}\mathbf{m} &  =-\gamma\mathbf{m}\times(\mathbf{H}+\mathbf{h})\cr &
+2\pi\gamma M_{s}\left(  m_{z}\mathbf{m}+n_{z}\mathbf{n}\right)
\times\mathbf{\hat{z}}\cr &  +\alpha_{0}(\mathbf{m}\times\partial
_{t}\mathbf{m}+\mathbf{n}\times\partial_{t}\mathbf{n})+\boldsymbol{\tau}%
_{m},\label{llgm}\\
\partial_{t}\mathbf{n} &  =-\gamma\mathbf{n}\times\left(  \mathbf{H}%
+\mathbf{h}+\frac{J_{ex}}{M_{s}d_{\mathrm{F}}}\mathbf{m}\right)  \cr & +2\pi\gamma
M_{s}\left(  n_{z}\mathbf{m}+m_{z}\mathbf{n}\right)  \times\mathbf{\hat{z}}\cr
&  +\alpha_{0}\left(  \mathbf{m}\times\partial_{t}\mathbf{n}+\mathbf{n}%
\times\partial_{t}\mathbf{m}\right)  +\boldsymbol{\tau}_{n},\label{llgn}%
\end{align}
where the spin-transfer torques $\boldsymbol{\tau}_{m}=(\boldsymbol{\tau}%
_{1}+\boldsymbol{\tau}_{2})/2$ and $\boldsymbol{\tau}_{n}=(\boldsymbol{\tau
}_{1}-\boldsymbol{\tau}_{2})/2$ become
\end{subequations}
\begin{subequations}
\begin{align}
\boldsymbol{\tau}_{m}/\alpha_{m} &  =\mathbf{m}\times\partial_{t}%
\mathbf{m}+\mathbf{n}\times\partial_{t}\mathbf{n}\cr &    +2\eta\frac
{\mathbf{m}\cdot(\mathbf{n}\times\partial_{t}\mathbf{n})}{1-\eta C}\mathbf{m}%
+2\eta\frac{\mathbf{n}\cdot(\mathbf{m}\times\partial_{t}\mathbf{m})}{1+\eta C}%
\mathbf{n},\label{sttm}\\
\boldsymbol{\tau}_{n}/\alpha_{n} &  =\mathbf{m}\times\partial_{t}%
\mathbf{n}+\mathbf{n}\times\partial_{t}\mathbf{m}\cr &    -2\eta\frac
{\mathbf{m}\cdot(\mathbf{n}\times\partial_{t}\mathbf{m})}{1+\eta C}\mathbf{m}%
-2\eta\frac{\mathbf{n}\cdot(\mathbf{m}\times\partial_{t}\mathbf{n})}{1-\eta C}%
\mathbf{n},\label{sttn}%
\end{align}
and $C=\mathbf{m}^{2}-\mathbf{n}^{2}$, while
\end{subequations}
\begin{subequations}
\begin{align}
\alpha_{m}&=\frac{\alpha_{1}g_{r}}{1+g_{r}\coth\left(  d_{\mathrm{N}}%
/2\lambda\right)  },\\
\alpha_{n}&=\frac{\alpha_{1}g_{r}}{1+g_{r}%
\tanh\left(  d_{\mathrm{N}}/2\lambda\right)  },\label{almn}%
\end{align}
\end{subequations}
with $\alpha_{1}=\gamma\hbar^{2}/(4e^{2}\lambda\rho M_{s}d_{\mathrm{F}})$.

\section{Calculation and results}

\label{result}


We consider the magnetization dynamics excited by linearly polarized
microwaves with a frequency $\omega$ and in-plane magnetic field
$\mathbf{h}(t)=(h_{x},h_{y},0)e^{i\omega t}$ that is much smaller than the
saturation magnetization. For small angle magnetization precession the total
magnetization and difference vector may be separated into a static equilibrium
and a dynamic component as $\mathbf{m}=\mathbf{m}_{0}+\delta\mathbf{m}$ and
$\mathbf{n}=\mathbf{n}_{0}+\delta\mathbf{n}$, respectively, where
$\mathbf{m}_{0}=(0,\sin\theta,0)$, $\mathbf{n}_{0}=(\cos\theta,0,0)$,
$C=-\cos2\theta$, and $\boldsymbol{s}=-\hat{\mathbf{z}}\sin2\theta$. The
equilibrium (zero torque) conditions $\mathbf{m}_{0}\times\mathbf{H}=0$ and
$\mathbf{n}_{0}\times(\mathbf{H}+J_{ex}/(M_{s}d_{\mathrm{F}})\mathbf{m}_{0})=0$ lead to
the relation
\begin{equation}
\sin\theta=H/H_{s},
\end{equation}
where $H_{s}=-J_{ex}/(M_{s}d_{\mathrm{F}})=|J_{ex}|/(M_{s}d_{\mathrm{F}})$ is the saturation
field. The LLG equations read
\begin{subequations}
\begin{align}
\partial_{t}\delta\mathbf{m}  &  =-\gamma\delta\mathbf{m}\times\mathbf{H}%
-\gamma\mathbf{m}_{0}\times\mathbf{h}\nonumber\\
&  +2\pi\gamma M_{s}\left(  \delta m_{z}\mathbf{m}_{0}+\delta n_{z}%
\mathbf{n}_{0}\right)  \times\hat{\mathbf{z}}\nonumber\\
&  +\alpha_{0}\left(  \mathbf{m}_{0}\times\partial_{t}\delta\mathbf{m}%
+\mathbf{n}_{0}\times\partial_{t}\delta\mathbf{n}\right)  +\delta
\boldsymbol{\tau}_{\mathrm{m}},\label{llgml}\\
\partial_{t}\delta\mathbf{n}  &  =-\gamma\delta\mathbf{n}\times\mathbf{H}%
-\gamma\mathbf{n}_{0}\times\mathbf{h}\nonumber\\
&  +2\pi\gamma M_{s}\left(  \delta n_{z}\mathbf{m}_{0}+\delta m_{z}%
\mathbf{n}_{0}\right)  \times\hat{\mathbf{z}}\nonumber\\
&  -\gamma H_{s}\left(  \mathbf{m}_{0}\times\delta\mathbf{n}-\mathbf{n}%
_{0}\times\delta\mathbf{m}\right) \nonumber\\
&  +\alpha_{0}\left(  \mathbf{m}_{0}\times\partial_{t}\delta\mathbf{n}%
+\mathbf{n}_{0}\times\partial_{t}\delta\mathbf{m}\right)  +\delta
\boldsymbol{\tau}_{\mathrm{n}}, \label{llgnl}%
\end{align}
with linearized spin-transfer torques
\end{subequations}
\begin{subequations}
\begin{align}
\delta\boldsymbol{\tau}_{m}/\alpha_{m}  &  =\mathbf{m}_{0}\times\partial
_{t}\delta\mathbf{m}+\mathbf{n}_{0}\times\partial_{t}\delta\mathbf{n}\cr &
\hskip-0.5cm-\frac{\eta\sin2\theta}{1+\eta\cos2\theta}\partial_{t}\delta n_{z}%
\mathbf{m}_{0}+\frac{\eta\sin2\theta}{1-\eta\cos2\theta}\partial_{t}\delta
m_{z}\mathbf{n}_{0},\label{sttml}\\
\delta\boldsymbol{\tau}_{n}/\alpha_{n}  &  =\mathbf{m}_{0}\times\partial
_{t}\delta\mathbf{n}+\mathbf{n}_{0}\times\partial_{t}\delta\mathbf{m}\cr &
\hskip-0.5cm+\frac{\eta\sin2\theta}{1-\eta\cos2\theta}\delta m_{z}\mathbf{m}_{0}%
-\frac{\eta\sin2\theta}{1+\eta\cos2\theta}\delta n_{z}\mathbf{n}_{0},
\label{sttnl}%
\end{align}
To leading order in the small precessing components $\delta\mathbf{m}$ and
$\delta\mathbf{n}$, the LLG equations in frequency space become
\end{subequations}
\begin{subequations}
\begin{align}
\delta{m_{x}}  &  =\gamma h_{x}\frac{\gamma\left(  H_{s}+4\pi M_{s}\right)
+i\omega\left(  \alpha_{0}+\frac{\alpha_{m}\left(  1+\eta\right)  }{1-\eta
\cos2\theta}\right)  }{\omega^{2}-\omega_{\mathrm{A}}^{2}-i\Delta_{\mathrm{A}%
}\omega}\sin^{2}\theta,\\
\delta{n_{y}}  &  =-\gamma h_{x}\frac{\gamma\left(  H_{s}+4\pi M_{s}\right)
+i\omega\left(  \alpha_{0}+\frac{\alpha_{n}\left(  1-\eta\right)  }{1-\eta
\cos2\theta}\right)  }{\omega^{2}-\omega_{\mathrm{A}}^{2}-i\Delta_{\mathrm{A}%
}\omega}\cos\theta\sin\theta,\\
\delta{m_{z}}  &  =-\gamma h_{x}\frac{i\omega}{\omega^{2}-\omega_{\mathrm{A}%
}^{2}-i\Delta_{\mathrm{A}}\omega}\sin\theta,
\end{align}%
\end{subequations}
\begin{subequations}
\begin{align}
\delta{n_{x}}  &  =-\gamma h_{y}\frac{4\pi\gamma M_{s}+i\omega\left(
\alpha_{0}+\frac{\alpha_{n}\left(  1-\eta\right)  }{1+\eta\cos2\theta}\right)
}{\omega^{2}-\omega_{\mathrm{O}}^{2}-i\Delta_{\mathrm{O}}\omega}\cos\theta
\sin\theta,\\
\delta{m_{y}}  &  =\gamma h_{y}\frac{4\pi\gamma M_{s}+i\omega\left(
\alpha_{0}+\frac{\alpha_{m}\left(  1+\eta\right)  }{1-\eta\cos2\theta}\right)
}{\omega^{2}-\omega_{\mathrm{O}}^{2}-i\Delta_{\mathrm{O}}\omega}\cos^{2}%
\theta,\\
\delta{n_{z}}  &  =\gamma h_{y}\frac{i\omega}{\omega^{2}-\omega_{\mathrm{O}%
}^{2}-i\Delta_{\mathrm{O}}\omega}\cos\theta.
\end{align}
The A modes ($\delta m_{x},\delta n_{y},\delta m_{z}$) are excited by $h_{x}$,
while the O modes ($\delta n_{x},\delta m_{y},\delta n_{z}$) couple to $h_{y}%
$. The poles in $\delta\mathbf{m}\left(  \omega\right)  \ $and $\delta
\mathbf{n}\left(  \omega\right)  $ define the resonance frequencies and
linewidths that do not depend on the magnetic field since we disregard
anisotropy and exchange-bias.

\subsection{Acoustic and Optical modes}

An antiferromagnetically exchange-coupled spin valves generally have
non-collinear magnetization configurations by the presence of external
magnetic fields. For $H<H_{s}$ ($0<\theta<\pi/2$), the acoustic mode:
\end{subequations}
\begin{align}
\omega_{\mathrm{{A}}}  &  =\gamma H\sqrt{1+(4\pi M_{s}/H_{s})},\label{w-AC}\\
\Delta_{\mathrm{{A}}}  &  =\alpha_{0}\gamma\left(  H_{s}+4\pi M_{s}+{H_{s}%
}\sin^{2}\theta\right) \nonumber\\
&  +\alpha_{m}\gamma\left(  H_{s}+4\pi M_{s}\right)  +\alpha_{\mathrm{{A}}%
}(\theta)\gamma{H_{s}}, \label{D-AC}%
\end{align}
and the optical mode:
\begin{align}
\omega_{\mathrm{{O}}}  &  =\gamma\sqrt{(4\pi M_{s}/H_{s})(H_{s}^{2}-H^{2}%
)},\label{w-OP}\\
\Delta_{\mathrm{{O}}}  &  =\alpha_{0}\gamma\left(  4\pi M_{s}+H_{s}\cos
^{2}\theta\right) \nonumber\\
&  +\alpha_{n}4\pi\gamma M_{s}+\alpha_{\mathrm{{O}}}(\theta)\gamma{H_{s}},
\label{D-OP}%
\end{align}
where
\begin{align}
\hskip-0.5cm\alpha_{\mathrm{A}}(\theta)  &  =\frac{\alpha_{1}g_{r}\sin
^{2}\theta}{1+g_{r}\tanh\left(  d_{\mathrm{N}}/2\lambda\right)  +2g_{r}%
\sin^{2}\theta/\sinh(d_{\mathrm{N}}/\lambda)},\\
\alpha_{\mathrm{O}}(\theta)  &  =\frac{\alpha_{1}g_{r}\cos^{2}\theta}%
{1+g_{r}\tanh\left(  d_{\mathrm{N}}/2\lambda\right)  +2g_{r}\cos^{2}%
\theta/\sinh(d_{\mathrm{N}}/\lambda)}.
\end{align}

\begin{figure}[ptb]
\includegraphics[width=0.45\textwidth,angle=0]{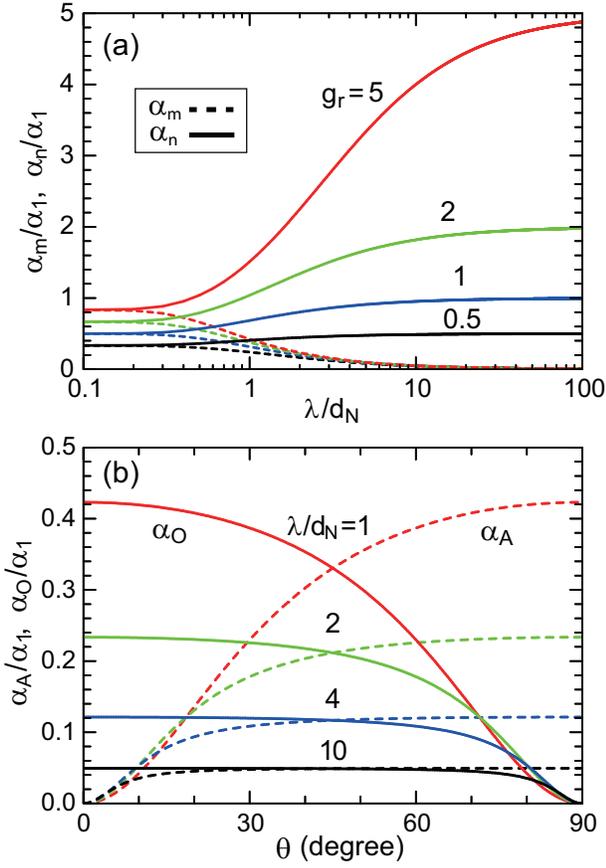}\caption{(a)
$\alpha_{m}$ (dashed line) and $\alpha_{n}$ (solid line) as a function of
$\lambda/d_{\mathrm{N}}$ for different values of the dimensionless mixing
conductance $g_{r}$. (b) $\alpha_{\mathrm{AC}}$ (dashed line) and
$\alpha_{\mathrm{OP}}$ (solid line) as a function of tilt angle $\theta$ for
$g_{r}=5$ and different values of $\lambda/d_{\mathrm{N}}$. }%
\label{fig:alpha}%
\end{figure}

The additional broadening in $\Delta_{\mathrm{A}}$ is proportional to
$\alpha_{m}$ and $\alpha_{\mathrm{A}}$ while that in $\Delta_{\mathrm{O}}$
scales with $\alpha_{n}$ and $\alpha_{\mathrm{O}}$ in.
Figure~\ref{fig:alpha}\thinspace\ (a) shows $\alpha_{m}$ and $\alpha_{n}$ as a
function of spacer layer thickness, indicating that $\alpha_{n}$ is always
larger than $\alpha_{m}$, and that $\alpha_{n}$ ($\alpha_{m}$) strongly
increases (decreases) with decreasing N layer thickness, especially for
$d_{\mathrm{N}}<\lambda$ and large $g_{r}$. Figure~\ref{fig:alpha}%
\thinspace(b) shows the dependence of $\alpha_{\mathrm{A}}$ and $\alpha
_{\mathrm{O}}$ on the tilted angle $\theta$ for different values of
$d_{\mathrm{N}}$. As $\theta$ increases, $\alpha_{\mathrm{A}}$ increases from
$0$ to $\alpha_{m}$ while $\alpha_{\mathrm{O}}$ decreases $\alpha_{m}$ to $0$.
The additional damping can be explained by the dynamic exchange. When two
magnetizations in spin valves precess in-phase, each magnet receives a spin
current that compensates the pumped one, thereby reducing the interface
damping. When the magnetizations precess out of phase, the $\pi$ phase
difference between both spin currents means that the moduli have to be added,
thereby enhancing the damping.

\begin{figure}[ptb]
\includegraphics[width=0.5\textwidth,angle=0]{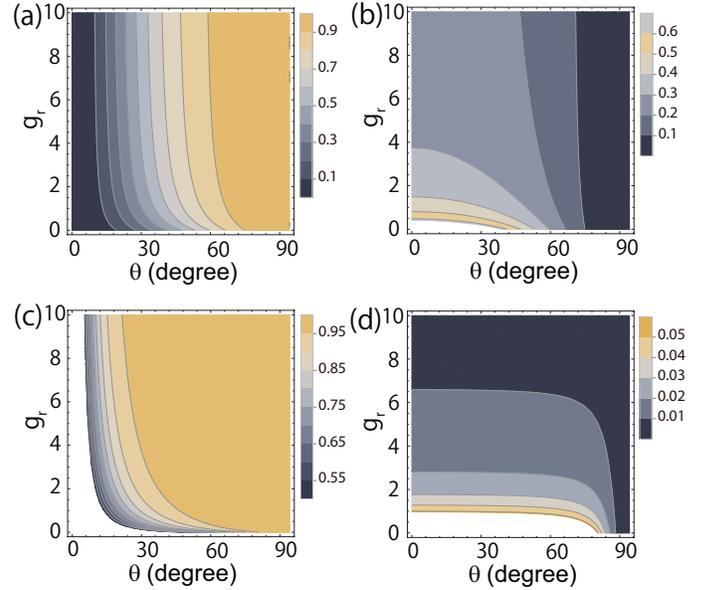}\caption{ (a,c)
$\alpha_{\mathrm{A}}(\theta)/\alpha_{m}$ and (b,d) $\alpha_{\mathrm{O}}%
(\theta)/\alpha_{n}$ as a function of $\theta$ and $g_{r}$ for different
values of $\lambda/d_{\mathrm{N}}$. (a,b) with $\lambda/d_{\mathrm{N}}=1$,
(c,d) with $\lambda/d_{\mathrm{N}}=10$ }%
\label{fig:alpha-ratio}%
\end{figure}

When the magnetizations are tilted by an angle $\theta$\ as sketched in
Fig.~\ref{sketch}, we predict an additional damping torque expressed by the
second terms of Eqs.~(\ref{interjs-a}) and (\ref{interjs-b}).
Figure~\ref{fig:alpha-ratio} shows the ratios $\alpha_{\mathrm{A}}%
(\theta)/\alpha_{m}$ and $\alpha_{\mathrm{O}}(\theta)/\alpha_{n}$ as a
function of $\theta$ and $g_{r}$ for different values of $\lambda
/d_{\mathrm{N}}$, thereby emphasizing the additional damping in the presence
of noncollinear magnetizations. In Fig.~\ref{fig:alpha-ratio}(a,b) with
$\lambda/d_{\mathrm{N}}=1$, i.e. for a spin-diffusive interlayer, the
additional damping of both A- and O-modes is significant in a large region of
parameter space. On the other hand, in Fig.~\ref{fig:alpha-ratio}(c,d) with
$\lambda/d_{\mathrm{N}}=10$, i.e. for an almost spin-ballistic interlayer, the
additional damping is more important for the A-mode, while the O-mode is
affected only close to the collinear magnetization. In the latter case the
intrinsic damping $\alpha_{0}$ dominates, however.


\subsection{In-phase and Out-of-phase modes}

When the applied magnetic field is larger than the saturation field ($H>H_{s}%
$), both magnetizations point in the $\hat{\mathbf{y}}$ direction, and the
$\delta\mathbf{m}$ (A) and $\delta\mathbf{n}$ (O) modes morph into in-phase
and $180^{\circ}$ out-of-phase (antiphase) oscillations of $\delta
\mathbf{m}_{1}$ and $\delta\mathbf{m}_{2}$, respectively. The resonance
frequency\cite{Zhang95} and linewidth of the in-phase mode for $H>H_{s}$
($\theta=\pi/2$) are
\begin{align}
\omega_{\mathrm{{A}}}  &  =\gamma\sqrt{H\left(  H+4\pi M_{s}\right)  },\\
\Delta_{\mathrm{{A}}}  &  =2(\alpha_{0}+\alpha_{m})\gamma\left(  H+2\pi
M_{s}\right)  , \label{flin}%
\end{align}
while those of the out-of-phase mode are
\begin{align}
\omega_{\mathrm{{O}}}  &  =\gamma\sqrt{\left(  H-H_{s}\right)  \left(
H-H_{s}+4\pi M_{s}\right)  },\\
\Delta_{\mathrm{{O}}}  &  =2\left(  \alpha_{0}+\alpha_{n}\right)
\gamma\left(  H-H_{s}+2\pi M_{s}\right)  .
\end{align}
\begin{figure}[ptb]
\includegraphics[width=0.45\textwidth,angle=0]{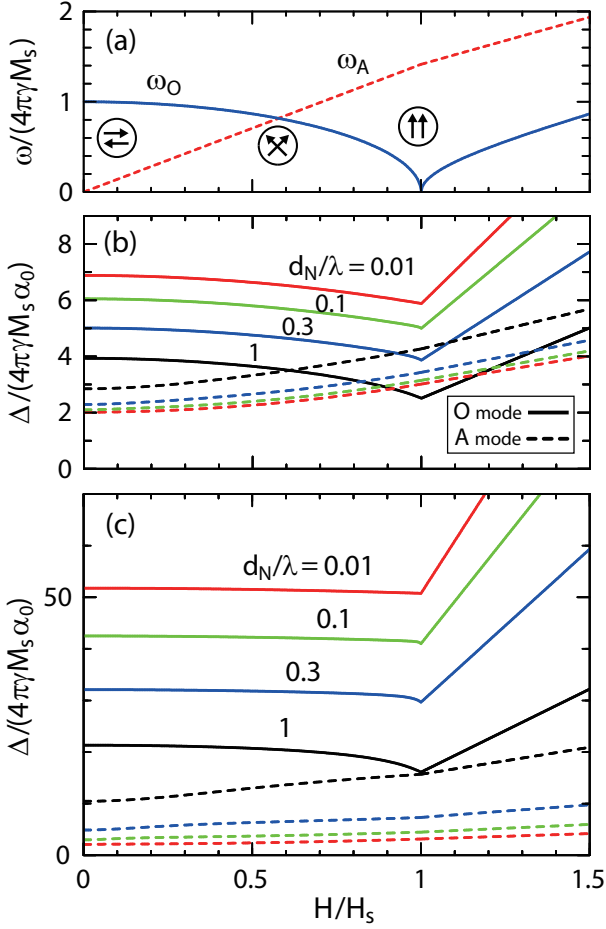}\caption{(a)
Resonance frequencies of the A and O modes as a function of magnetic field for
$H_{s}/(4\pi M_{s})=1$. (b), (c) Linewidths of the A (dashed line) and the O
(solid line) modes for $H_{s}/(4\pi M_{s})=1$, $g_{r}=5$, and different values
of $d_{\mathrm{N}}/\lambda$. (b) $\alpha_{1}/\alpha_{0}=1$ and (c) $\alpha
_{1}/\alpha_{0}=10$. }%
\label{fig:LW}%
\end{figure}

Figure~\ref{fig:LW}\thinspace(a) shows the calculated resonance frequencies of
the A and O modes as a function of an applied magnetic field $H$ while
\ref{fig:LW}\thinspace(b) displays the linewidths for $\alpha_{1}/\alpha
_{0}=1$, which is representative for ferromagnetic metals, such as permalloy
(Py) with an intrinsic magnetic damping of the order of $\alpha_{0}=0.01$ and
a comparable additional damping $\alpha_{1}$ due to spin pumping. A value
$g_{r}=4/5$ corresponds to $\lambda=20/200$\textit{$\,\mathrm{nm}$, }%
$\rho=10/2.5\,\mathrm{\mu\Omega cm}$\textit{ }for
N=Ru/Cu,\cite{Eid02,Yakata06}$G_{r}=2/1\times10^{15}\mathrm{\Omega}%
^{-1}\mathrm{m}^{-2}$ for the N$|$Co(Py) interface\cite{JiaPRB11,Xia02}, and
$d_{\mathrm{F}}=1\,\mathrm{nm}$, for example. The colors in the figure represent
different relative layer thicknesses $d_{\mathrm{N}}/\lambda$. The linewidth
of the A mode in Fig.~\ref{fig:LW}\thinspace(b) increases with increasing $H$,
while that of the O mode starts to decrease until a minimum at the saturation
field $H=H_{s}$. Figure~\ref{fig:LW}\thinspace(c) shows the linewidths for
$\alpha_{1}/\alpha_{0}=10$, which describes ferromagnetic materials with low
intrinsic damping, such as Heusler alloys \cite{Mizukami09} and magnetic
garnets.\cite{Coey10} In this case, the linewidth of the O mode is much larger
than that of the A mode, especially for small $d_{\mathrm{N}}/\lambda$.

In the limit of $d_{\mathrm{N}}/\lambda\rightarrow0$ is easily established
experimentally. The expressions of the linewidth in Eqs.~(\ref{D-AC}) and
(\ref{D-OP}) are then greatly simplified to $\Delta_{\mathrm{A}}=\gamma
(H_{s}+4\pi M_{s}+H_{s}\sin^{2}\theta)\alpha_{0}$ and $\Delta_{\mathrm{O}%
}=\gamma\left(  4\pi M_{s}+H_{s}\cos^{2}\theta\right)  \alpha_{0}+(4\pi\gamma
M_{s})g_{r}\alpha_{1}$, and $\Delta_{\mathrm{A}}\ll\Delta_{\mathrm{O}}$ when
$g_{r}\alpha_{1}\gg\alpha_{0}$. The additional damping, Eq.~(\ref{almn})
reduces to $\alpha_{m}\rightarrow0$ and $\alpha_{n}\rightarrow2[\gamma
\hbar/(4\pi M_{s}d_{\mathrm{F}})(h/e^{2})G_{r}]$ when the magnetizations are
collinear and in the ballistic spin transport limit.\cite{Tserkovnyak05} In
contrast to the acoustic mode, the dynamic exchange interaction enhances
damping of the optical mode. $\Delta_{\mathrm{O}}\gg\Delta_{\mathrm{A}}$ has
been observed in Py$|$Ru$|$Py trilayer spin valves \cite{Belmeguenai07} and
Co$|$Cu multilayers \cite{Tanaka14}, consistent with the present results.

For spin valves with ferromagnetic metals, the interface backflow spin-current
[(\ref{jsb})] reads
$\mathbf{J}_{j}^{\mathrm{B}}=(G_{r}/e)\left[  \xi_{\mathrm{F}}\left(  \mathbf{m}%
_{j}\cdot\boldsymbol{\mu}_{sj}\right)  \mathbf{m}_{j}-\boldsymbol{\mu}%
_{sj}\right]  ,$
where $\xi_{\mathrm{F}}=1-(G/2G_{r})(1-p^{2})(1-\eta_{\mathrm{F}})$ $(0\leq\xi_{\mathrm{F}}\leq1)$, $G$
is the N%
$\vert$%
F interface conductance per unit area, and $p$ the conductance spin
polarization.\cite{Jiao13} Here the spin diffusion efficiency is
\begin{equation}
\frac{1}{\eta_{\mathrm{F}}}=1+\frac{\sigma_{\mathrm{F}}}{G\lambda_{\mathrm{F}}}\frac{\tanh(d_{\mathrm{F}}%
/\lambda_{\mathrm{F}})}{\cosh(d_{\mathrm{F}}/\lambda_{\mathrm{F}})},
\end{equation}
where $\sigma_{\mathrm{F}}$, $\lambda_{\mathrm{F}}$, and $d_{F}$ are the conductivity, the
spin-flip diffusion length, and the layer thickness of the ferromagnets,
respectively. For the material parameters of a typical ferromagnet with
$d_{\mathrm{F}}=1\ \mathrm{nm}$, the resistivity $\rho_{\mathrm{F}}=10\ \mathrm{\mu\Omega cm}$,
$G=2G_{r}=10^{15}\ \Omega^{-1}\mathrm{m}^{-2}$, $\lambda_{\mathrm{F}}=10\ \mathrm{nm}$,
and $p=0.7$, $\xi_{\mathrm{F}}=0.95$, which justifies disregarding this contribution
from the outset.
\begin{figure}[ptb]t
\includegraphics[width=0.45\textwidth,angle=0]{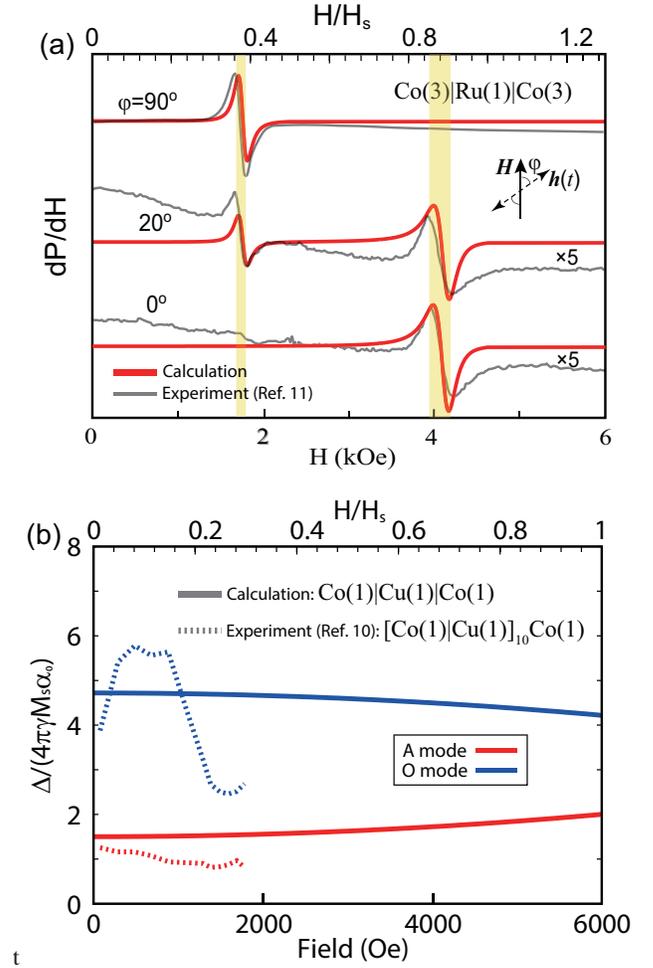}
\caption{ (a) Derivative of the microwave absorption spectrum $dP/dH$ at
frequency $\omega/(2\pi)=9.22\ \mathrm{GHz}$ for different angles $\varphi$
between the microwave field and the external magnetic field for $H_{s}/(4\pi
M_{s})=0.5$, $\omega/(4\pi\gamma M_{s})=0.35$, $d_{\mathrm{N}}/\lambda=0.1$,
$d_{\mathrm{F}}/\lambda=0.3$ $\alpha_{0}=\alpha_{1}=0.02$, and $g_{r}=4$. The
experimental data have been adopted from Ref.~\onlinecite{Zhang94}. (b)
Computed linewidths of the A and O modes of a Co$|$Cu$|$Co spin valve (dashed
line) compared with experiments on a Co$|$Cu multilayer (solid
line).\cite{Tanaka14} }%
\label{fig:DPO}%
\end{figure}

\section{Comparison with experiments}

\label{experiment}

FMR experiments yield the resonant absorption spectra of a microwave field of
a ferromagnet. The microwave absorption power $P=2\left\langle \mathbf{h}%
(t)\cdot\partial_{t}\mathbf{m}(t)\right\rangle $ becomes in our model
\begin{align}
P  &  =\frac{1}{4}\frac{\gamma^{2}M_{s}(H_{s}+4\pi M_{s})\Delta_{\mathrm{A}}%
}{(\omega-\omega_{\mathrm{A}})^{2}+(\Delta_{\mathrm{A}}/2)^{2}}h_{x}^{2}%
{\sin^{2}}\theta\nonumber\\
&  +\frac{1}{4}\frac{\gamma^{2}M_{s}(4\pi M_{s})\Delta_{\mathrm{O}}}%
{(\omega-\omega_{\mathrm{O}})^{2}+(\Delta_{\mathrm{O}}/2)^{2}}h_{y}^{2}%
{\cos^{2}}\theta.
\end{align}
$P$ depends sensitively on the character of the resonance, the polarization of
the microwave, and the strength of the applied magnetic field. In
Figure~\ref{fig:DPO}\thinspace(a) we plot the normalized derivative of the
microwave absorption spectra $dP/(P_{0}dH)$ at different angles $\varphi$
between the microwave field $\mathbf{h}(t)$ and the external magnetic field
$\mathbf{H}$, where $P_{0}=\gamma M_{s}h^{2}$ and $\mathbf{h}(t)=h(\sin
\varphi,\cos\varphi,0)e^{i\omega t}$. Here we use the experimental values
$H_{s}=5$\thinspace kOe, $4\pi M_{s}=10$\thinspace kOe, $d_{\mathrm{N}}%
=1$\thinspace nm, $d_{\mathrm{F}}=3$\thinspace nm, and microwave frequency
$\omega/(2\pi)=9.22$\thinspace GHz as found for a symmetric Co$|$Ru$|$Co
trilayer.\cite{Zhang94} $\lambda=20\,\mathrm{nm}$ for Ru, $\alpha_{0}%
=\alpha_{1}=0.02$, and $g_{r}=4$ is adopted (corresponding to $G_{r}%
=2\times10^{15}\mathrm{\Omega}^{-1}\mathrm{m}^{-2})$\textit{.} When
$\mathbf{h}(t)$ is perpendicular to $\mathbf{H}$ ($\varphi=90^{\circ}$), only
the A mode is excited by the transverse ($\delta m_{x},\delta m_{z}$)
component. When $\mathbf{h}(t)$ is parallel to $\mathbf{H}$ ($\varphi
=0^{\circ}$), the O mode couples to the microwave field by the longitudinal
$\delta m_{y}$ component. For intermediate angles ($\varphi=20^{\circ}$), both
modes are excited at resonance. We observe that the optical mode signal is
broader than the acoustic one, as calculated. The theoretical resonance
linewidths of the A and O modes as well as the absorption power as a function
of microwave polarization reproduce the experimental results for
Co(3.2\thinspace nm)$|$Ru(0.95\thinspace nm)$|$Co(3.2\thinspace nm)
well.\cite{Zhang94}

Figure~\ref{fig:DPO}\thinspace(b) shows the calculated linewidths of A and O
modes as a function of an applied magnetic field for a Co(1\thinspace
nm)$|$Cu(1\thinspace nm)$|$Co(1\thinspace nm) spin valve. The experimental
values $\lambda=200\,\mathrm{nm}$ and $\rho=2.5\,\mathrm{\mu\Omega cm}$ for
Cu, $\alpha_{0}=0.01$ and $4\pi M_{s}=15$\thinspace kOe for Co, and $g_{r}=5$
(corresponding to $G_{r}=10^{15}\Omega^{-1}\mathrm{m}^{-2}$) for the interface
have been adopted.\cite{Xia02} We partially reproduce the experimental data
for magnetic multilayers; for the weak-field broadenings of the observed
linewidths agreement is even quantitative. The remaining discrepancies in the
applied magnetic field dependence might reflect
exchange-dipolar\cite{Hans14SW} and/or multilayer\cite{Tanaka14} spin waves
beyond our spin valve model in the macrospin approximation.

\section{Conclusions}

\label{conclusion}

In summary, we modelled the magnetization dynamics in antiferromagnetically
exchange-coupled spin valves as a model for synthetic antiferromagnets. We
derive the Landau-Lifshitz-Gilbert equations for the coupled magnetizations
including the spin transfer torques by spin pumping based on the spin
diffusion model with quantum mechanical boundary conditions. We obtain
analytic expressions for the linewidths of magnetic resonance modes for
magnetizations canted by applied magnetic fields and achieve good agreement
with experiments. We find that the linewidths strongly depend on the type of
resonance mode (acoustic and optical) as well as the strength of magnetic
fields. The magnetic resonance spectra reveal complex magnetization dynamics
far beyond a simple precession even in the linear response regime. Our
calculated results compare favorably with experiments, thereby proving the
importance of dynamic spin currents in these devices. Our model calculation
paves the way for the theoretical design of synthetic AFM material that is
expected to play a role in next-generation spin-based data-storage and
information technologies.

\section{Acknowledgments}

The authors thanks K. Tanaka, T. Moriyama, T. Ono, T. Yamamoto, T. Seki, and
K. Takanashi for valuable discussions and collaborations. This work was
supported by Grants-in-Aid for Scientific Research (Grant Nos. 22540346,
25247056, 25220910, 268063) from the JSPS, FOM (Stichting voor Fundamenteel
Onderzoek der Materie), the ICC-IMR, EU-FET Grant InSpin 612759, and DFG
Priority Programme 1538 \textquotedblleft Spin-Caloric Transport" (BA 2954/2).

\appendix

\section{Coupled Landau-Lifshitz-Gilbert equations in noncollinear spin
valves}

\label{Appendix}

Both magnets and interfaces in our NM$|$F$|$NM spin valves are assumed to be
identical with saturation magnetization $M_{s}$ and $G_{r}$ the real part of
the spin-mixing conductance per unit area (vanishing imaginary part). When
both magnetizations are allowed to precess as sketched in Fig.~\ref{sketch}
(a), the LLG equations expanded to include additional spin-pump and
spin-transfer torques read
\begin{widetext}
\begin{align}
\frac{\partial\mathbf{m}_{i}}{\partial t}  &  =-\gamma\mathbf{m}_{i}\times\mathbf{H_{\mathrm{eff}i}%
}+\alpha_{0i}\mathbf{m}_{i}\times\frac{\partial\mathbf{m}_{i}}{\partial t}\nonumber\\
&+\alpha_{\mathrm{SP}i}\left[  \mathbf{m}_{i}\times\frac{\partial\mathbf{m}_{i}}{\partial t}-\eta\mathbf{m}_{j}\times\frac{\partial\mathbf{m}_{j}}{\partial t}+\eta
\left(\mathbf{m}_{i}\cdot\mathbf{m}_{j}\times\frac{\partial\mathbf{m}_{j}}{\partial t}\right)\mathbf{m}_{i}\right]+\alpha_{\mathrm{SP}i}^{\mathrm{nc}}(\varphi)
\mathbf{m}_{i}\times\left(\mathbf{m}_{i}\times\mathbf{m}_{j}\right)
,\label{llg-nc}\\
&\alpha_{\mathrm{SP}i}^{\mathrm{nc}}(\varphi)=\frac{\alpha_{\mathrm{SP}i}%
\eta^{2}}{1-\eta^{2}(\mathbf{m}_{i}\cdot\mathbf{m}_{j})^{2}}\left[  \mathbf{m}_{j}\cdot\mathbf{m}_{i}\times\frac{\partial\mathbf{m}_{i}}{\partial t}+\eta
\left(\mathbf{m}_{i}\cdot\mathbf{m}_{j}\times\frac{\partial\mathbf{m}_{j}}{\partial t}\right)(\mathbf{m}_{j}\cdot\mathbf{m}_{i})\right],\label{al-nc}
\end{align}
\end{widetext}
where $\gamma$ and $\alpha_{0i}$ are the gyromagnetic ratio and the Gilbert
damping constant of the isolated ferromagnetic films labeled by $i$ and
thickness $d_{\mathrm{F}i}$. Asymmetric spin valves due to the thickness difference
$d_{\mathrm{F}i}$ suppress the cancellation of mutual spin-pump in A-mode, which may be
advantage to detect both modes in the experiment. The effective magnetic field
\begin{equation}
\mathbf{H}_{\mathrm{eff}i}=\mathbf{H}_{i}+\mathbf{h}(t)+\mathbf{H}%
_{\mathrm{di}i}(t)+\mathbf{H}_{\mathrm{ex}j}(t)
\end{equation}
consists of the Zeeman field $\mathbf{H}_{i}$, a microwave field
$\mathbf{h}(t)$, the dynamic demagnetization field $\mathbf{H}_{\mathrm{di}%
i}(t)$, and interlayer exchange field $\mathbf{H}_{\mathrm{ex}j}(t)$. The
Gilbert damping torque parameterized by $\alpha_{0i}$ governs the relaxation
towards an equilibrium direction. The third term in Eq.~(\ref{llg-nc})
represents the mutual spin pumping-induced damping-like torques in terms of
damping parameter

\begin{equation}
\alpha_{\mathrm{SP}i}=\frac{\gamma\hbar^{2}G_{r}}{2e^{2}M_{s}d_{\mathrm{F}i}}\frac{\eta
S}{1-\eta^{2}},
\end{equation}
where%
\begin{equation}
\eta=\frac{g_{r}}{\sinh(d_{\mathrm{N}}/\lambda)+g_{r}\cosh(d_{\mathrm{N}%
}/\lambda)}%
\end{equation}
and $g_{r}=2\lambda\rho G_{r}$ is dimensionless. The fourth term in Eq.
(\ref{llg-nc}) is the damping Eq.~(\ref{al-nc}) that depends on the relative
angle $\varphi$ between the magnetizations. When $\mathbf{m}_{j}$ is fixed
 along the $\mathbf{H}_{i}$ direction, i.e. a spin-sink limit, Eq.~(\ref{llg-nc}) reduces to
the dynamic stiffness in spin valves without an electrical bias.\cite{Stiff}

When the magnetizations are noncollinear as in Fig.~\ref{sketch}, we have to
take into account the additional damping torques described by the second terms
in Eqs.~(\ref{interjs-a}) and (\ref{interjs-b}$,$). In the ballistic limit
$d_{\mathrm{N}}/\lambda\rightarrow0$ and collinear magnetizations,
Eq.~(\ref{llg-nc}) reduces to the well known LLG equation with dynamic
exchange interaction.\cite{Tserkovnyak05,Heinrich03,Timopheev14}


\begin{thebibliography}{99}                                                                                               %


\bibitem {Marti14}X. Marti, I. Fina, C. Frontera, Jian Liu, P.Wadley, Q. He,
R. J. Paull, J. D. Clarkson, J. Kudrnovsk\'{y}, I. Turek, J. Kune\v{s}, D. Yi,
J-H. Chu, C. T. Nelson, L. You, E. Arenholz, S. Salahuddin, J. Fontcuberta, T.
Jungwirth, and R. Ramesh, Nat. Mater. \textbf{13}, 367 (2014).

\bibitem {Park11}B. G. Park, J. Wunderlich, X. Mart\'{\i}, V. Hol\'{y}, Y.
Kurosaki, M. Yamada, H. Yamamoto, A. Nishide, J. Hayakawa, H. Takahashi, A. B.
Shick, and T. Jungwirth, Nat. Mater. \textbf{10}, 347 (2011).

\bibitem {Haney07}P. M. Haney, D. Waldron, R. A. Duine, A. S. N\'{u}{n}ez, H.
Guo, and A. H. MacDonald, Phys. Rev. B \textbf{75}, 174428 (2007).

\bibitem {Sharma07}Z.Wei, A. Sharma, A. S. Nunez, P. M. Haney, R. A. Duine, J.
Bass, A. H. MacDonald, and M. Tsoi, Phys. Rev. Lett. \textbf{98}, 116603 (2007).

\bibitem {Urazhdin07}S. Urazhdin and N. Anthony, Phys. Rev. Lett. \textbf{99},
046602 (2007).

\bibitem {Wang08}Y. Xu, S. Wang, and K. Xia, Phys. Rev. Lett. \textbf{100},
226602 (2008).

\bibitem {Hals11}K. M. D. Hals, Y. Tserkovnyak, and A. Brataas, Phys. Rev.
Lett. \textbf{106}, 107206 (2011).

\bibitem {Gomonay12}H. V. Gomonay, R. V. Kunitsyn, and V. M. Loktev, Phys.
Rev. B \textbf{85}, 134446 (2012).

\bibitem {Hahn14}C. Hahn, G. de Loubens, V. V. Naletov, J. B. Youssef, O.
Klein, and M. Viret, Eur. Phys. Lett. \textbf{108}, 57005 (2014).

\bibitem {Wang14}H. Wang, C. Du, P. C. Hammel, and F. Yang, Phys. Rev. Lett.
\textbf{113}, 097202 (2014).

\bibitem {Moriyama14}T. Moriyama, M. Nagata, K. Tanaka, K. Kim, H. Almasi,
W.Wang, and T. Ono, ArXiv e-prints (2014), arXiv:1411.4100.

\bibitem {Duine11}R. Duine, Nat Mater \textbf{10}, 344 (2011).

\bibitem {MacDonald11}A. H. MacDonald and M. Tsoi, Phil. Trans. Royal Soc. A
\textbf{369}, 3098 (2011).

\bibitem {Kimel05}A. V. Kimel, A. Kirilyuk, P. A. Usachev, R. V. Pisarev, A.
M. Balbashov, and Th. Rasing, Nature (London) \textbf{435}, 655 (2005).

\bibitem {Satoh10}T. Satoh, S.-J. Cho, R. Iida, T. Shimura, K. Kuroda, H.
Ueda, Y. Ueda, B. A. Ivanov, F. Nori, and M. Fiebig, Phys. Rev. Lett.
\textbf{105}, 077402 (2010).

\bibitem {Wienholdt12}S. Wienholdt, D. Hinzke, and U. Nowak, Phys. Rev. Lett.
\textbf{108}, 247207 (2012).

\bibitem {Grunberg86}P. Gr\"{u}nberg, R. Schreiber, Y. Pang, M. B. Brodsky,
and H. Sowers, Phys. Rev. Lett. \textbf{57}, 2442 (1986).

\bibitem {Grunberg07}P. Gr\"{u}nberg, D. E. B\"{u}rgler, H. Dassow, A. D.
Rata, C. M. Schneider, Acta Materialia \textbf{55}, 1171 (2007).

\bibitem {Parkin91}S. S. P. Parkin, X. Jiang, C. Kaiser, A. Panchula, K.
Roche, and M. Samant, Proc. IEEE \textbf{91}, 661 (2003).

\bibitem {Baibich88}M. N. Baibich, J. M. Broto, A. Fert, F. Nguyen Van Dau, F.
Petroff, P. Etienne, G. Creuzet, A. Friederich, and J. Chazelas, Phys. Rev.
Lett. \textbf{61}, 2472 (1988).

\bibitem {Binasch89}G. Binasch, P. Gr\"{u}nberg, F. Saurenbach, and W. Zinn,
Phys. Rev. B \textbf{39}, 4828 (1989).

\bibitem {Brataas12}A. Brataas, A. D. Kent, and H. Ohno, Nat. Mater.
\textbf{11}, 372 (2012).

\bibitem {Chen10}E. Chen, D. Apalkov, Z. Diao, A. Driskill-Smith, D. Druist,
D. Lottis, V. Nikitin, X. Tang, S. Watts, S. Wang, S. Wolf, A. W. Ghosh, J.
Lu, S. J. Poon, M. Stan, W. Butler, S. Gupta, C. K. A. Mewes, T. Mewes, and P.
Visscher, IEEE Trans. Magn. \textbf{46}, 1873 (2010).

\bibitem {Ralph08}D. C. Ralph and M. D. Stiles, J. Magn. Magn. Mater. 320,
1190 (2008).

\bibitem {Carey08}M. J. Carey, N. Smith, S. Maat, and J. R. Childress, Appl.
Phys. Lett. \textbf{93}, 102509 (2008).

\bibitem {Tserkovnyak02}Y. Tserkovnyak, A. Brataas, G. E. W. Bauer, Phys. Rev.
Lett \textbf{88}, 117601 (2002).

\bibitem {Tserkovnyak05}Y. Tserkovnyak, A. Brataas, G. E. W. Bauer, and B. I.
Halperin, Rev. Mod. Phys. \textbf{77}, 1375 (2005).

\bibitem {Heinrich03}B. Heinrich, Y. Tserkovnyak, G. Woltersdorf, A. Brataas,
R. Urban, and G. E. W. Bauer, Phys. Rev. Lett. \textbf{90}, 187601 (2003).

\bibitem {Takahashi14}S. Takahashi, Appl. Phys. Lett. \textbf{104}, 052407 (2014).

\bibitem {Tanaka14}K. Tanaka, T. Moriyama, M. Nagata, T. Seki, K. Takanashi,
S. Takahashi, and T. Ono Appl. Phys. Express \textbf{7}, 063010 (2014).

\bibitem {Zhang94}Z. Zhang, L. Zhou, P. E. Wigen, and K. Ounadjela, Phys. Rev.
Lett. \textbf{73}, 336 (1994).

\bibitem {Belmeguenai07}M. Belmeguenai, T. Martin, G. Woltersdorf, M. Maier,
and G. Bayreuther, Phys. Rev. B \textbf{76}, 104414 (2007).

\bibitem {Hans14}H. Skarsv\aa g, G. E. W. Bauer, and A. Brataas, Phys. Rev. B
\textbf{90}, 054401 (2014)

\bibitem {Stamps94}R. L. Stamps, Phys. Rev. B \textbf{49}, 339 (1994).

\bibitem {Demokritov01}S. O. Demokritov, B. Hillebrands, and A. N. Slavin,
Phys. Rep. \textbf{348}, 441 (2001).

\bibitem {Cochran90}J. F. Cochran, J. Rudd, W. B. Muir, B. Heinrich, and Z.
Celinski, Phys. Rev. B \textbf{42}, 508 (1990).

\bibitem {Kuanr02}B. K. Kuanr, M. Buchmeier, D. E. B\"{u}rgler, and P.
Gr\"{u}nberg, J. Appl. Phys. \textbf{91}, 7209 (2002).

\bibitem {Timopheev14}A. A. Timopheev, Yu. G. Pogorelov, S. Cardoso, P. P.
Freitas, G. N. Kakazei, and N. A. Sobolev, Phys. Rev. B \textbf{89}, 144410 (2014).

\bibitem {Taniguchi07}T. Taniguchi and H. Imamura Phys. Rev. B \textbf{76},
092402 (2007).

\bibitem {Zhou13}Y. Zhou, J. Xiao, G.E.W. Bauer, and F. C. Zhang, Phys. Rev. B
\textbf{87}, 020409(R) (2013).

\bibitem {Layadi97}A. Layadi and J. O. Artman, J. Magn. Magn. Mate.
\textbf{176}, 175 (1997).

\bibitem {Kuanr03}B. K. Kuanr, M. Buchmeier, R. R. Gareev, D. E. B\"{u}rgler,
R. Schreiber, and P. Gr\"{u}nberg, J. Appl. Phys. \textbf{93}, 3427 (2003).

\bibitem {Evans14}R. F. L. Evans, T. A. Ostler, R. W. Chantrell, I. Radu, and
T. Rasing, Appl. Phys. Lett. \textbf{104}, 082410 (2014).

\bibitem {Gonzalez-Chavez13}D. E. Gonzalez-Chavez, R. Dutra, W. O. Rosa, T. L.
Marcondes, A. Mello, and R. L. Sommer, Phys. Rev. B \textbf{88}, 104431 (2013).

\bibitem {Liu14}X. M. Liu, Hoa T. Nguyen, J. Ding, M. G. Cottam, and A. O.
Adeyeye, Phys. Rev. B \textbf{90}, 064428 (2014).

\bibitem {Krebs90}J. J. Krebs, P. Lubitz, A. Chaiken, and G. A. Prinz, J.
Appl. Phys. \textbf{67}, 5920 (1990).

\bibitem {Chiba14}T. Chiba, G. E. W. Bauer, and S. Takahashi, Phys. Rev. Appl.
\textbf{2}, 034003 (2014)

\bibitem {Chiba15}T. Chiba, M. Schreier, G. E. W. Bauer, and S. Takahashi, J.
Appl. Phys. \textbf{117}, 17C715 (2015).

\bibitem {Hans14SW}H. Skarsv\aa g, Andr\'{e} Kapelrud, and A. Brataas, Phys.
Rev. B \textbf{90}, 094418 (2014).

\bibitem {Yuan14}Z. Yuan, K. M. D. Hals, Y. Liu, A. A.
Starikov, A. Brataas, and P. J. Kelly, Phys. Rev. Lett. \textbf{113}, 266603 (2014).

\bibitem {Jiao13}H. Jiao and G. E.W. Bauer, Phys. Rev. Lett. \textbf{110},
217602 (2013).

\bibitem {Jia11}X. Jia, K. Liu, K. Xia, and G. E. W. Bauer, Europhys. Lett.
\textbf{96}, 17005 (2011).

\bibitem {Zhang95}Z. Zhang and P. E. Wigen, \textit{High frequency processes
in magnetic materials}, edited by G. Srinivasan and A. N. Slavin (World
Scientific, Singapore, 1995).

\bibitem {Eid02}K. Eid, R. Fonck, M. AlHaj Darwish, W. P. Pratt Jr., and J.
Bass, J. Appl. Phys. \textbf{91}, 8102 (2002).

\bibitem {Yakata06}S. Yakata, Y. Ando, T. Miyazaki, and S. Mizukami, Jpn.. J.
Appl. Phys. \textbf{45}, 5A (2006).

\bibitem {JiaPRB11}X. Jia, Y. Li, K. Xia, and G. E. W. Bauer, Phys. Rev. B
\textbf{84}, 134403 (2011).

\bibitem {Xia02}K. Xia, P. J. Kelly, G. E. W. Bauer, A. Brataas, and I. Turek,
Phys. Rev. B \textbf{65}, 220401 (2002).

\bibitem {Mizukami09}S. Mizukami, D. Watanabe, M. Oogane, Y. Ando, Y. Miura,
M. Shirai, and T. Miyazaki, J. Appl. Phys. \textbf{105}, 07D306 (2009).

\bibitem {Coey10}J. M. D. Coey, \textit{Magnetism and magnetic materials}
(Cambridge University Press, Cambridge, 2010).

\bibitem {Stiff}Y. Tserkovnyak, A. Brataas, G. E. W. Bauer, Phys. Rev. B
\textbf{67}(R), 140404(2003)
\end{thebibliography}
\end{document}